\documentclass[aps,prb,twocolumn,floatfix]{revtex4}
\usepackage{amsmath,amssymb,dsfont,graphicx,psfrag}
\usepackage{color,epsfig,rotate}

\usepackage[normalem]{ulem} 

\begin{document}

\title{N{\'e}el to spin-Peierls transition in a quasi-1D Heisenberg model coupled to bond phonons}

\author{Jason Cornelius Pillay, Keola Wierschem and Pinaki Sengupta} 
\address{School of Physical and Mathematical Sciences, Nanyang Technological University, Singapore 637371}

\begin{abstract}
The zero and finite temperature spin-Peierls transitions in a quasi-one-dimensional spin-$1\over 2$ 
Heisenberg model coupled to adiabatic bond phonons is investigated using the Stochastic Series 
Expansion (SSE) Quantum Monte Carlo (QMC) method. The quantum phase transition from a 
gapless N\'eel state to a spin-gapped Peierls state is studied in the parameter space spanned by 
spatial anisotropy, inter-chain coupling strength and spin-lattice coupling strength. It is found that 
for any finite inter-chain coupling, the transition to a dimerized Peierls ground state only occurs 
when the spin-lattice coupling exceeds a finite, non-zero critical value. This is in contrast to the 
pure 1D model (zero inter-chain coupling), where adiabatic/classical phonons lead to a dimerized 
ground state for any non-zero spin-phonon interaction. The phase diagram in the parameter space 
shows that for a strong inter-chain coupling, the relation between the inter-chain coupling and the 
critical value of the spin-phonon interaction is linear whereas for weak inter-chain coupling, this 
behavior is found to have a natural logarithm-like relation. No region was found to have a long 
range magnetic order and dimerization occurring simultaneously. Instead, the N\'eel state order 
vanishes simultaneously with the setting in of the spin-Peierls state. For the thermal phase transition, 
a continuous heat capacity with a peak at the critical temperature, $T_{c}$, shows a second order 
phase transition. The variation of the equilibrium bond length distortion, $\delta_{eq}$, with 
temperature showed a power law relation which decayed to zero as the temperature was increased 
to $T_{c}$, indicating a continuous transition from the dimerized phase to a paramagnetic phase with 
uniform bond length and zero antiferromagnetic susceptibility.
\end{abstract}

\maketitle

\section{Introduction}
A spin-{$1\over 2$} Heisenberg chain coupled to an elastic lattice is unstable towards dimerization. 
The cost in elastic energy due to a distortion, $\delta$, of the lattice ($\sim \delta^2$) is smaller
than the accompanying   gain in the magnetic energy ($\sim \delta^{4/3}$). This causes the ground 
state to be stabilized for a lattice with non-zero dimerization\cite{pytte,cross} with a finite spin gap. 
The transition to such a dimerized phase is known as the spin-Peierls (SP) transition, analogous to 
the conventional Peierls transition in one-dimensional (1D) metals. In the adiabatic limit, any arbitrarily 
small coupling to an elastic lattice leads to a dimerized ground state in a spin chain. This is contrary 
to quantum phonons where quantum lattice fluctuations destroy the bond distortions for small spin-phonon 
couplings and/or large bare phonon frequencies\cite{aws-sp,bursill} -  the transition to a Peierls state 
occurs only when the spin-phonon coupling exceeds a finite, non-zero critical value (that depends on 
the bare phonon frequency). The discovery\cite{hase} of the quasi-1D inorganic spin-Peierls compound 
CuGeO$_3$ led to a resurgence in the study of the spin-Peierls transition in low-dimensional spin 
models. The properties of CuGeO$_3$ have been widely studied within the framework of a 1D 
spin-${1\over 2}$ Heisenberg model coupled to phonons, with an additional frustrated next-nearest-neighbor 
interaction.\cite{riera}

Further studies have shown that inter-chain coupling in  CuGeO$_3$  is not negligible and is estimated 
to be $J_\perp/J\approx 0.1$.\cite{nishi} As such, a more realistic modeling of the real material requires 
the study of the spin-Peierls transition in 2D or quasi-1D systems. In contrast to spin chains, the ground 
state of the $S=1/2$ Heisenberg model in 2D (in the absence of spin-phonon coupling) has long range 
antiferromagnetic (N{\'e}el) order. It is generally believed that even for adiabatic phonons, the spin-phonon 
coupling has to exceed some non-zero critical value for the ground state to develop a dimerized pattern with 
a spin gap. What is the nature of the transition? Is there a region in the phase space where the ground state 
has co-existing dimerization and long range antiferromagnetic order? The existence of several different possible 
dimerization patterns in 2D means that, in principle, different dimerization patterns can be stabilized for different 
values of the parameters. Finally, unlike 1D, the Peierls phase in 2D extends to finite temperatures.

The spatially isotropic 2D spin-$1\over 2$ Heisenberg model with static dimerization patterns has been studied 
by several authors.\cite{if1,if2,katoh,koga,al-omari,sirker} In these previous works, the energetically most favored 
dimerization pattern was predicted by comparing the ground state energies for different patterns. Using this same 
method, the 2D tight binding model with bond-distortions\cite{ono} and the 2D Peierls-Hubbard 
model\cite{tang,sumit,zhang} have also been studied. In the limit of large on-site repulsion, $U$, the Hubbard model 
at half-filling reduces to the Heisenberg model, thus, results from the Peierls-Hubbard model (in the limit of large $U$) 
should be applicable to the present discussion. However, there is no consensus among the different studies as to the 
nature of the dimerization pattern in the ground state. For the Peierls-Hubbard model at half-filling, Tang and 
Hirsch\cite{tang} find a plaquette-like distortion to be energetically favored in the limit of large $U$. On the other hand, 
Mazumdar\cite{sumit} has argued that the minimum energy ground state has a ``stairlike'' dimerization pattern -- 
corresponding to a wave vector $\bf{Q}=(\pi,\pi)$. Zhang and Prelov\v{s}ek\cite{zhang} agree with a dimerization pattern 
with $\bf{Q}=(\pi,\pi)$, but conclude that the ground state has dimerization only along one of the axes (staggered dimerized 
chains -- similar to the pattern considered here). For the Heisenberg model with static dimerization, Al-Omari\cite{al-omari} 
has concluded that the ground state energy is minimized for a plaquette-like distortion of the lattice which agrees with the
 conclusion of Tang and Hirsch. On the other hand, Sirker {\it et.al.}\cite{sirker} find that linear spin wave theory (LSWT)
 predicts a stairlike dimerization pattern to be most favored, in agreement with Mazumdar. Using LSWT, Sirker {\it et.al.} 
also find finite regions in the parameter space with co-existing long range magnetic order and non-zero dimerization. 
However, as pointed out by the authors, results obtained from LSWT are not reliable at large values of dimerization. 
The effects of inter-chain coupling was considered early on by Inagaki and Fukuyama\cite{if1,if2} who studied a 
quasi-1D system of coupled spin-${1\over 2}$ Heisenberg chains with a fixed dimerization pattern corresponding to a 
wave vector ${\bf Q}=(\pi,0)$. By treating the inter-chain coupling in a mean-field theory, they were able to map out the 
ground state phase diagram\cite{if1} and study the finite temperature transition\cite{if2}. Later, Katoh and Imada\cite{katoh} 
studied in detail the nature of the transition. More recently, the effects of impurities have been studied in this model (once 
again with a mean-field treatment of the inter-chain coupling) which revealed a region of co-existing Peierls and 
antiferromagnetic orders.\cite{khomskii,saito,affleck,dobry,melin1,melin2} In addition to this, the quasi-1D $XY$ model 
with a ${\bf Q}=(\pi,\pi)$ dimerization pattern has recently been studied\cite{ji,yuan} by an extension of the Jordan-Wigner 
transformation in 2D.\cite{wang1,wang2,azzouz} The effects of quantum phonons in the isotropic 2D model have also been 
studied,\cite{low} where the authors find that there is no evidence of a transition to the Peierls state for a wide range of values 
of the bare phonon frequency and the spin-phonon coupling. This is consistent with a previous finding\cite{rts} for the same model 
that the spin wave spectrum along the Brillouin zone boundary is qualitatively similar to that for the pure Heisenberg model.

As can be seen from the discussion above, it can be difficult to uniquely determine the optimal dimerization pattern in a given model. Further, when comparing to experimental results, it is not clear whether the optimal dimerization pattern of a model will be robust to the presence of additional interactions that may occur in the material. Thus, rather than determining the optimum dimerization pattern for a system of weakly coupled Heisenberg chains coupled to bond phonons, we instead choose a fixed dimerization pattern. A natural choice is the stairlike $\bf{Q}=(\pi,\pi)$ dimerization pattern that has been experimentally observed in CuGeO$_3$.\cite{tranquada}

In this study, we aim to investigate the SP transition in a spin-${1\over 2}$ quasi-1D Heisenberg 
antiferromagnet coupled to static $\bf{Q}=(\pi,\pi)$ bond phonons. By comparing the elastic energy 
cost and magnetic energy gain associated with a finite bond distortion at fixed  interaction strengths,
the complete ground state phase diagram is mapped out in the parameter space of inter-chain coupling
and the strength of spin-lattice interaction. The nature of the N{\' e}el-SP quantum phase transition and
the evolution of magnetic and Peierls order across the transition are investigated in detail. In the second
part of the study, the nature of the thermal transition out of the SP state is investigated --  including determination
of the universality class --  by studying the variation of bond-length distortion and specific heat across the 
transition.

The rest of the paper is organized as follows. In Section II, the model Hamiltonian and the Stochastic 
Series Expansion (SSE) QMC method used to study it are introduced. The results of the simulations 
are presented in Section III. Section IV concludes with a summary of the results.

\section{Model and Simulation techniques}

The Stochastic Series Expansion (SSE) Quantum Monte Carlo (QMC) method was used to study a 
quasi-1D Heisenberg model with spin-phonon coupling. The model is given by the Hamiltonian
\begin{eqnarray}
H &=& J\sum_{i,j}(1+\lambda u_{i,j}){\bf S}_{i,j}\cdot{\bf S}_{i+1,j} + {1\over 2}K\sum_{i,j}u_{i,j}^2 \nonumber \\
  & & +J_\perp\sum_{i,j}{\bf S}_{i,j}\cdot{\bf S}_{i,j+1},
\end{eqnarray}
where $J_\perp$ is the inter-chain coupling, $\lambda$ is the strength of the spin-phonon coupling 
(restricted to be only along the chains), $u_{i,j}$'s are the distortions of the bond lengths and $K$ 
is the elastic constant for the distortions. Following the experimentally observed\cite{tranquada} 
dimerization pattern in CuGeO$_3$, the bond length distortions are chosen to be of the form
\[
u_{i,j}=(-1)^{i+j}\delta
\]
This amounts to choosing a fixed dimerization pattern along the chains corresponding to the 
wave-vector ${\bf Q}=(\pi,\pi)$. The bond distortions can be rescaled by the spin-phonon coupling 
strength, $\lambda$, thereby reducing the Hamiltonian to
\begin{eqnarray}
H &=& \sum_{i,j}(1+(-1)^{i+j}\delta ){\bf S}_{i,j}\cdot{\bf S}_{i+1,j} + N\delta^2/2\zeta \nonumber \\
  & & +\alpha\sum_{i,j}{\bf S}_{i,j}\cdot{\bf S}_{i,j+1},
\label{eqn:H}
\end{eqnarray}
where $\zeta={\frac{\lambda^2J}{K}}$, $\alpha=J_\perp/J$ and $N$ is the size of the lattice. 
The static approximation for the displacements makes the computational task easier for the
ground state determination since one needs to minimize only the total energy. The following 
strategy is adopted. The simulations are carried out for the spin variables for several different 
$\{\alpha,\delta\}$ parameter sets. This produces the spin energy of the system as a function 
of $\delta$ for a fixed $\alpha$. Next the elastic energy with a particular $\zeta$ is added and 
the total energy is minimized to obtain the value of the ground state distortion for the given set 
of parameters $\{\alpha,\zeta\}$. This is repeated for different sets of $\{\alpha,\zeta\}$ to obtain 
the ground state phase diagram in the parameter space spanned by $\alpha$ and $\zeta$.

The above approach fails for the finite temperature studies, where one needs to minimize the free energy (the entropic contribution is non-zero at finite T). Instead, both the equilibrium bond distortion and spin configurations are dynamically determined by Monte Carlo updates.
The stochastic series expansion (SSE) QMC method has been used to sample spin configurations in the present model. The SSE method \cite{aws1,aws2,sseloop} is a finite-temperature quantum Monte Carlo method based on importance sampling of the diagonal elements of the Taylor expansion of $e^{-\beta H}$, where $\beta$ is the inverse temperature $\beta=J/T$. Ground state expectation values can be obtained using sufficiently large values of $\beta$, and there are no approximations beyond statistical errors. The use of loop updates\cite{sseloop,dirloops} makes it possible to explore the spin configuration space of the Hamiltonian (\ref{eqn:H}) in an efficient manner. 

To sample the equilibrium bond distortion, we use a straightforward implementation of the Metropolis Monte Carlo algorithm.\cite{metropolis}
In order to simplify the bond update, we separate the elastic term $\epsilon(\delta)=N\delta^2/2\zeta$ from the spin-dependent portion of the Hamiltonian. By the linear property of the trace, the partition function then becomes $e^{-\beta\epsilon(\delta)}Tr(e^{-\beta[H-\epsilon(\delta)]})$. Treating the spin-dependent portion of the Hamiltonian with the SSE method,~\cite{aws1} configuration weights thus become $e^{-\beta\epsilon(\delta)}W(\alpha,\delta, S)$, where $W$ is the weight of the spin configuration $S$ generated by the spin updates (note that $W$ also depends on the Hamiltonian parameters $\alpha$ and $\delta$). During the bond update, a new value of $\delta$ is randomly chosen from a discritezed grid: $\delta'=\delta\pm\Delta$, where $\Delta=0.01$ has been used in the present implementation. The proposed move is then accepted with probability
\begin{equation}
MIN\left[1,\frac{e^{-\beta\epsilon(\delta')}}{e^{-\beta\epsilon(\delta)}}\frac{W(\alpha,\delta', S)}{W(\alpha,\delta, S)}\right].
\end{equation}
The weights $W$ are easily calculated within the SSE framework, so we do not include them here.

For a static dimerization pattern, the above update simply amounts to sampling $\delta$, the bond distortion parameter. A single Monte Carlo move thus changes the bond distortion of the entire lattice simultaneously. This acts as a global update, and no critical slowing down is expected. Examination of autocorrelation times confirms this intuitive picture. On the other hand, if every bond is updated separately (crucial for determining the optimal dimerization pattern for a model), such a strategy would lead to long autocorrelation times. Instead, one needs to use more sophisticated approach\cite{miyashita} in such situations.

\begin{figure}
\centering
\includegraphics[clip,trim=0cm 1cm 5cm 11cm,width=0.95\linewidth]{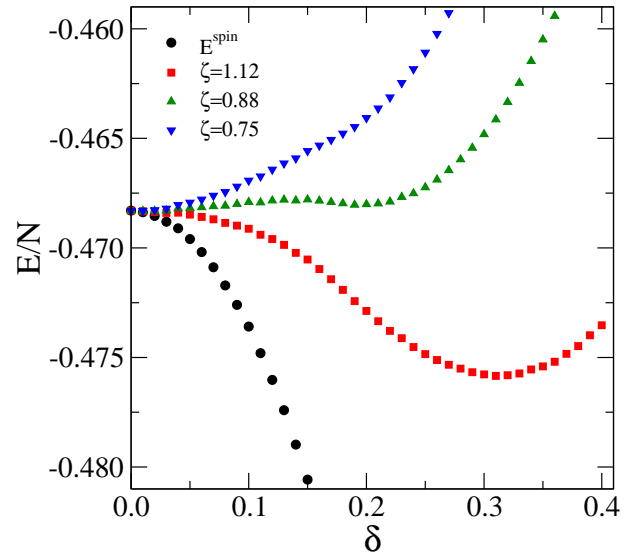}
\caption{The magnetic (spin) and the total ground state energy per site as a function of
the bond distortion for 3 representative values of the elastic constant, $\zeta$,
and fixed $\alpha$=0.25. The system size is $N$=256x32.} 
\label{fig:etot}
\end{figure}

\section{Results}

We begin with the determination of the nature of the ground state for different parameter
regimes. For a system of weakly coupled Heisenberg chains, it was shown that the 
estimates for various observables for a spatially anisotropic system depend 
non-monotonically on the system size for square ($L_x = L_y$) geometry.\cite{multichain} 
One has to go to rectangular ($L_x \neq L_y$) geometries to obtain monotonic 
behavior of the numerical results for extrapolating to the thermodynamic limit.
This is essentially due to a finite-size gap of the coupled chains that scales as 
$~1/L_x$. Only when this gap is smaller than the energy scale of coupling between 
chains can we begin to approach the thermodynamic limit. Thus, while square lattices 
eventually converge to the correct thermodynamic values in the limit of infinite system 
size, they do so much more slowly than appropriately chosen rectangular lattices.
Similar effects are expected in the present model for $\alpha \ll 1$. Hence 
rectangular lattices with the aspect ratio $L_x = 4L_y$ have been studied, with 
$L_x=16-512$. An inverse temperature of $\beta=8L_x$ was found to be sufficient 
for the observables to have converged to their ground state values. The inter-chain 
coupling is varied over $0.006 \le \alpha < 1$, concentrating in the regime $\alpha < 0.1$.

\begin{figure}
\centering
\includegraphics[clip,trim=0cm 0cm 0cm 0cm,width=0.95\linewidth]{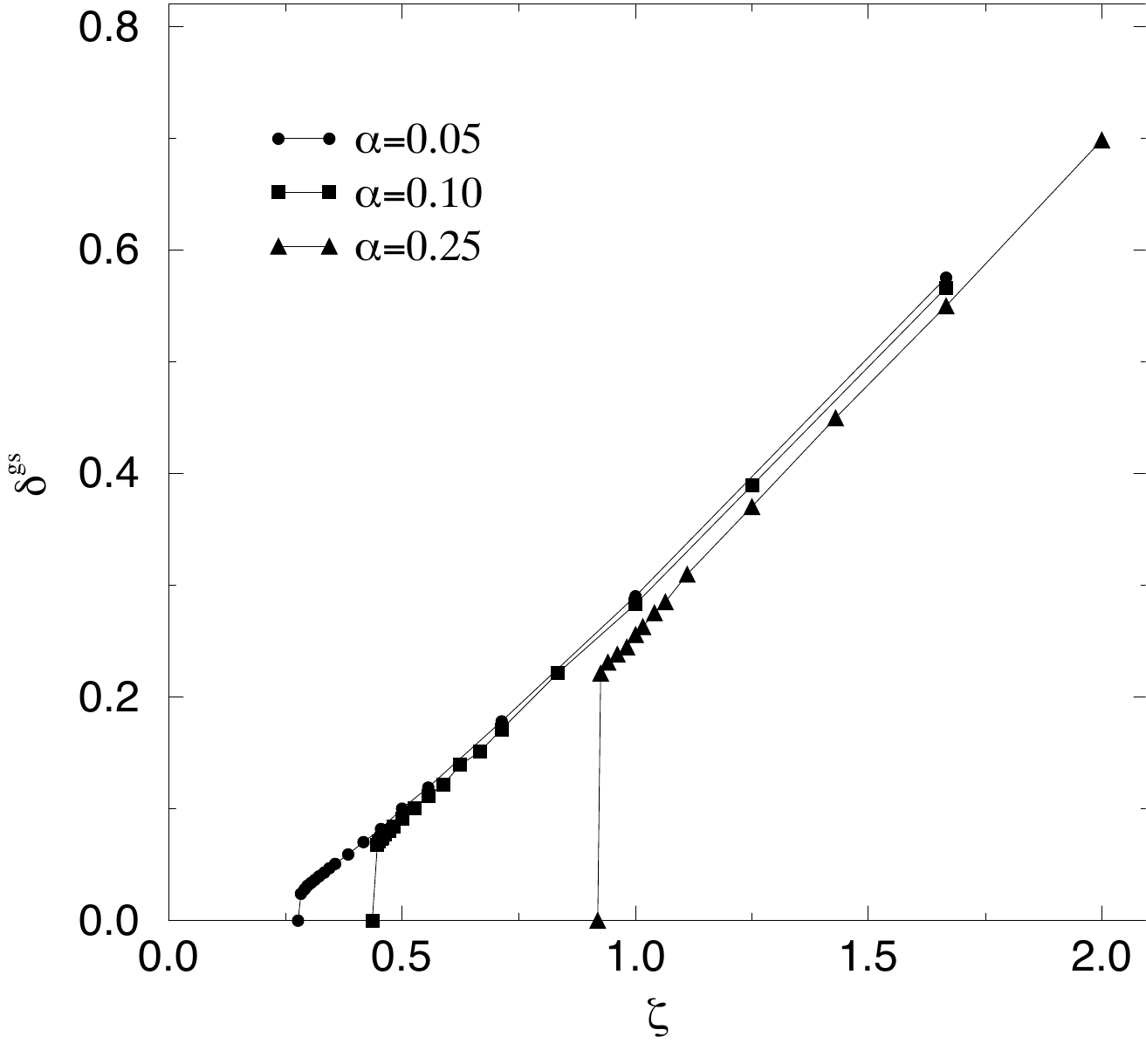}
\caption{The equilibrium ground state bond distortion as a function of the
spin-lattice coupling for different values of the inter-chain coupling.} 
\label{fig:deq}
\end{figure}

As discussed earlier, for the determination of the ground state phases, the magnetic
energy is calculated for a range of values of the static bond distortion and the elastic
energy is added subsequently to determine the total energy. For small values of $\delta$, 
the leading order finite-size correction to the ground state energy is seen to be 
$\sim 1/L_x^3$ -- similar to that observed for the pure 2D Heisenberg model.\cite{aws2} 
On the other hand, for large values of $\delta$, when the ground state is expected to be
in the Spin-Peierls phase, the energy scales exponentially with system-size. Close to the 
critical point, extrapolation to the thermodynamic limit becomes difficult due to cross-over 
effects. Instead, the data from the largest system size studied have been used to map out 
the phase diagram. Fortunately, the data for the largest system sizes studied are found to 
be well converged -- the fractional difference in the energy for the two largest system sizes 
studied is $\sim 10^{-5}$. This observed convergence allows for reliable estimation of ground 
state properties in the thermodynamic limit based on the data from the largest system size -- 
any finite-size effects on such estimates are expected to be small.

The strategy implemented to extract the ground state bond distortion is qualitatively 
demonstrated in fig.\ref{fig:etot}. The total ground state energy is obtained by adding the 
elastic energy contribution to the spin part of the energy obtained from the simulations. 
The plot shows the spin part of the energy, as well as the total ground state energy as a 
function of the bond length distortion, $\delta$, for three representative values of the elastic 
energy constant, $\zeta$, at a fixed value of the inter-chain coupling ($\alpha$=0.25). 
For large $\zeta$, the total energy is minimum for a non-zero value of the bond length 
distortion, $\delta$, which implies a SP ground state. On the other hand, for small 
$\zeta$, a uniform ground state with $\delta$=0 is energetically favored. The behavior 
of the total energy near the critical $\zeta$ is also shown. The ground state distortion 
in bond length is obtained by numerically differentiating the total energy data and 
solving for $\frac{\partial {\mbox E^{tot}}}{\partial\delta}|_{\delta^{gs}}=0$. In principle, 
one can also fit a polynomial to the QMC data to get ${\mbox E^{spin}}(\delta)$ and 
add to it the elastic energy term to get ${\mbox E^{tot}}(\delta)$. The ground state 
distortion can then be obtained as a continuous function of $\delta$ by solving 
$\frac{\partial {\mbox E^{tot}}}{\partial\delta}|_{\delta^{gs}}=0$. However, in practice, 
the numerical minimization is found to be more reliable because of the uncertainty 
in the order of the polynomial fit.

Fig.\ref{fig:deq} shows the equilibrium distortion in the bond lengths in the ground 
state of the system as a function of the elastic energy parameter, $\zeta$, at fixed 
values of $\alpha$, obtained as described above. For small values of $\zeta$, the 
tendency towards dimerization is suppressed, and a uniform (N{\'e}el ordered) 
ground state with uniform bond lengths is stabilized. As $\zeta$ is increased 
above a critical value, $\zeta_c$, there is a discontinuous (first order) quantum 
phase transition to a ground state with a finite, non-zero dimerization. For 
$\zeta > \zeta_c$, the equilibrium ground state distortion increases monotonically 
with $\zeta$. 

For uncoupled chains with only Heisenberg interaction, the ground state has
no true long range magnetic ordering -- it is a critical state with algebraically 
decaying spin-spin correlations. An infinitesimally small  $\zeta$ is sufficient to
destroy the algebraic correlation and the ground state is dimerized for all 
non-zero spin-phonon coupling. A finite inter-chain coupling establishes true 
long range N{\' e}el ordering and consequently a finite  $\zeta_c$ required for 
a transition to a dimerized ground state. The critical $\zeta_c$ increases with 
increasing inter-chain coupling.

\begin{figure}
\centering
\includegraphics[clip,trim=0cm 1cm 5cm 11cm,width=0.95\linewidth]{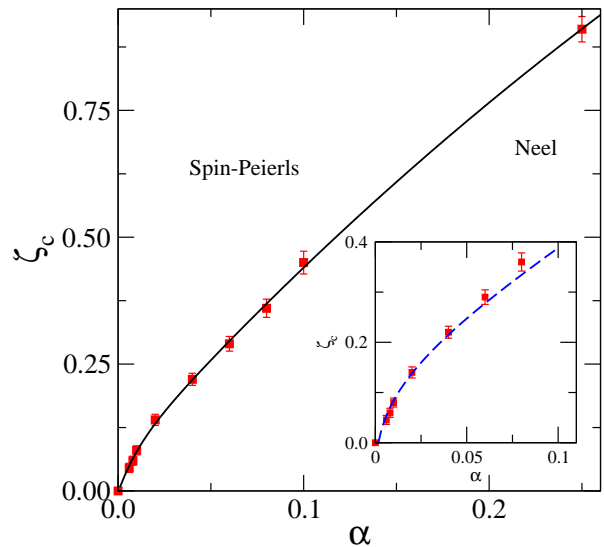}
\caption{The ground-state phase diagram in the parameter space of the
inter-chain coupling and the spin-lattice coupling strength. The inset shows 
the results for $\alpha<0.1$ and the logarithmic dependence of $\zeta_c$
as $\alpha \rightarrow 0$.} 
\label{fig:phase}
\end{figure}

The results from fig.\ref{fig:deq} are summarized in fig.\ref{fig:phase}, which 
shows the ground state phase diagram for the Hamiltonian (\ref{eqn:H}) in 
the phase space spanned by the parameters $\zeta$ and $\alpha$. For 
small $\zeta$ and/or large $\alpha$, the ground state of the system is 
N{\'e}el ordered with zero bond distortion, while for large $\zeta$ and/or 
small $\alpha$, the ground state is dimerized with a finite spin gap. The 
critical coupling strength goes to zero as $\zeta_c \sim 1/\mbox{ln}\alpha$ 
as $\alpha \rightarrow 0$. This is consistent with a similar behavior of the
N{\' e}el temperature, $T_N$, for a system of coupled Heisenberg 
chains.~\cite{yasuda} Since both $T_N$ and $\zeta_c$ are approximate 
measures of the energy required to destroy 
the N{\' e}el ordering, such a similarity in asymptotic behavior is expected.
Note that while one could argue that the energy scale of the SP phase 
(i.e. the spin gap) may also exhibit a dependence on $\alpha$, it is clear 
from Fig.\ref{fig:deq} that this dependence is very small. Thus, to a first 
approximation, we do not expect significant corrections to the 
$1/\mbox{ln}\alpha$ behavior of the phase boundary described above.

\begin{figure}
\centering
\includegraphics[clip,trim=0cm 0cm 0cm 0cm,width=0.95\linewidth]{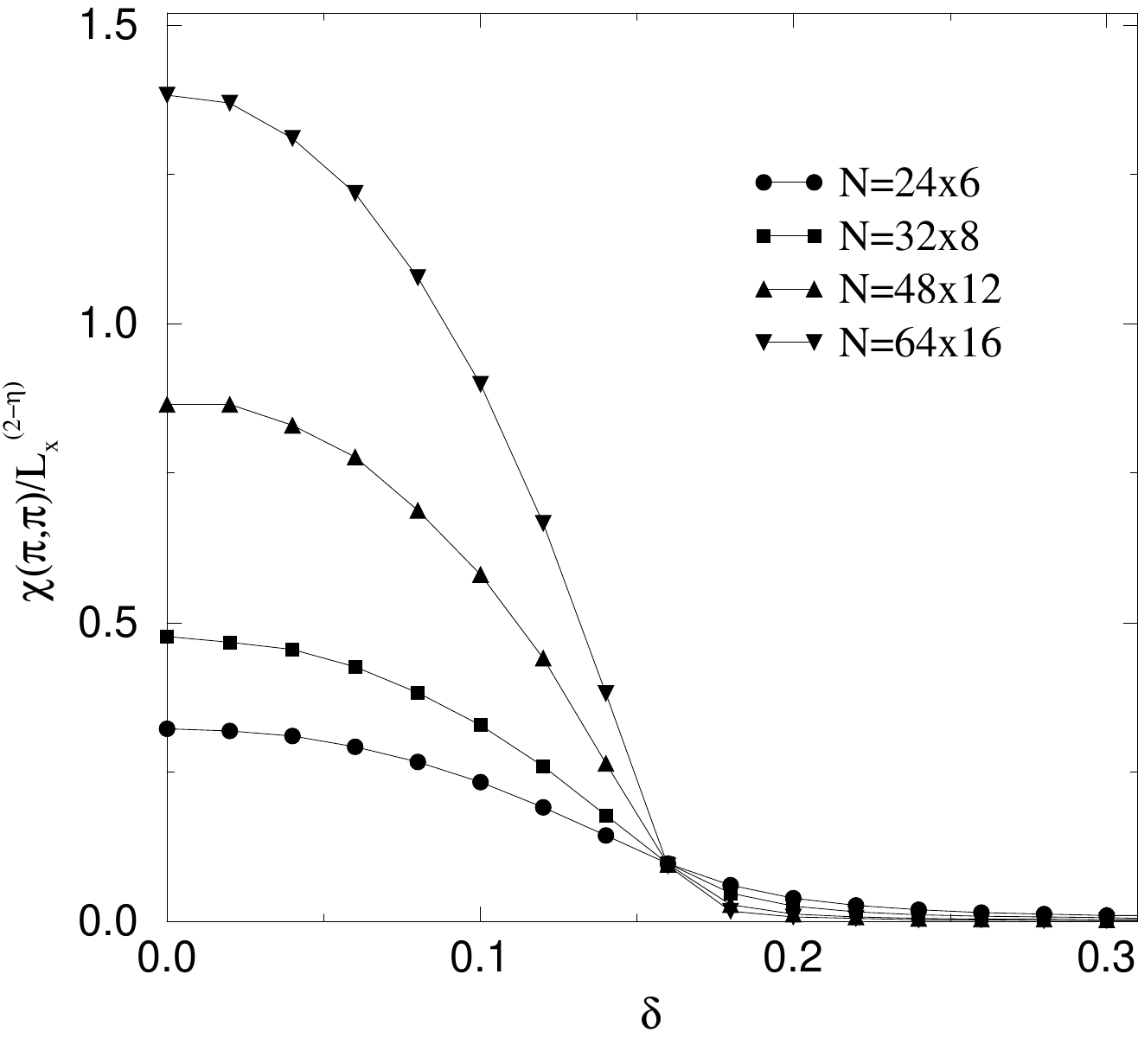}
\caption{The ground-state static staggered magnetic susceptibility 
as a function of bond distortion for a fixed value of the inter-chain
coupling, $\alpha=0.25$.}
\label{fig:xs}
\end{figure}

Next we turn to the determination of magnetic properties in the ground state 
phase with zero bond-distortion. This also raises the interesting possibility 
of having a region in the ($\zeta,\alpha$) parameter space where the ground 
state has co-existing N{\'e}el order and non-zero dimerization. Such co-existence 
has been shown to exist in the presence of doping.\cite{khomskii,saito,affleck,dobry,melin1,melin2} 
For this purpose, the static spin susceptibility, defined as
\begin{equation}
\chi({\bf q})={1\over N}\sum_{i,j}e^{i{\bf q}\cdot ({\bf r}_i-{\bf r}_j)}\int^{\beta}_0d\tau\langle S^z_j(\tau)S^z_i(0) \rangle,
\end{equation}
has been studied for the spin part of the Hamiltonian (\ref{eqn:H}) (without the elastic energy term):
\begin{eqnarray}
H &=& \sum_{i,j}(1+(-1)^{i+j}\delta ){\bf S}_{i,j}\cdot{\bf S}_{i+1,j} + \nonumber \\
  & & +\alpha\sum_{i,j}{\bf S}_{i,j}\cdot{\bf S}_{i,j+1}.
\label{eqn:Hspin}
\end{eqnarray}
If the ground state has long range antiferromagnetic order, the staggered 
(${\bf Q}=(\pi,\pi)$) susceptibility scaled by the system size ($\chi(\pi,\pi)/N$), 
for a finite system will increase with increasing system size, diverging in the 
thermodynamic limit. On the other hand, if the ground state has a finite spin 
gap, $\chi(\pi,\pi)/N$ will vanish in the limit of infinite system size. This qualitative 
criterion can be expressed in a more quantitative manner by noting that the 
ground state of the above Hamiltonian undergoes a continuous transition 
from a N{\'e}el ordered state with long range antiferromagnetic order to a 
spin-gapped, dimerized phase with no long range magnetic order as $\delta$ 
is increased beyond a finite, non-zero critical value, $\delta^*$, that depends 
on the inter-chain coupling $\alpha$. The transition belongs to the universality 
class of the 3D Heisenberg model\cite{chn}. Finite-size scaling\cite{barber} 
predicts that for such a transition, the finite-size susceptibility at the critical 
$\delta$ scales with the system size as
\[
\chi(L_x)\sim L_x^{2-\eta},
\]
for a rectangular lattice of dimension $N=L_x$x$L_y$. This implies that on 
a plot of $\chi(\pi,\pi)/L_x^{2-\eta}$, the curves for different system sizes will 
intersect at the critical $\delta$. The value of the critical exponent $\eta$ is 
known to a high degree of accuracy ($\eta \approx 0.037$).\cite{campostrini} 
Fig. \ref{fig:xs} shows  $\chi(\pi,\pi)/L_x^{2-\eta}$ as a function of $\delta$ for 
a fixed value of $\alpha = 0.25$ for several different system sizes. For small 
$\delta$, the scaled susceptibility increases with increasing system size, 
indicating the presence of long range magnetic order. For larger values of 
$\delta$, the scaled susceptibility goes to zero with increasing system size, 
signaling the opening up of a spin gap. From the data, the critical value of 
$\delta$ is estimated to be $\delta^* \approx 0.16$. This value of the bond 
length distortion is less than the jump in $\delta$ at the transition point, 
$\zeta_c$. This means that for $\zeta > \zeta_c$, the ground state is dimerized 
with $\delta^{gs} > \delta^*$ and has no long range magnetic order. On the 
other hand, for $\zeta < \zeta_c$, a uniform ground state is energetically 
favored that has zero dimerization ($\delta^{gs}=0$), and long range magnetic 
order (the N{\'e}el state). For no values of $\zeta$ is a ground state with 
$0 < \delta^{gs} < \delta^*$ ever favored energetically. Hence the transition 
to the dimerized phase is accompanied by the simultaneous vanishing of magnetic 
order and there is no region of co-existing dimerization and magnetic order. 
This is true for the present model. Whether it is possible to have ground 
states with co-existing magnetic order and dimerization in other models 
(with different dimerization patterns) remains to be seen.

\begin{figure}
\centering
\includegraphics[clip,trim=0cm 0cm 0cm 7cm,width=0.95\linewidth]{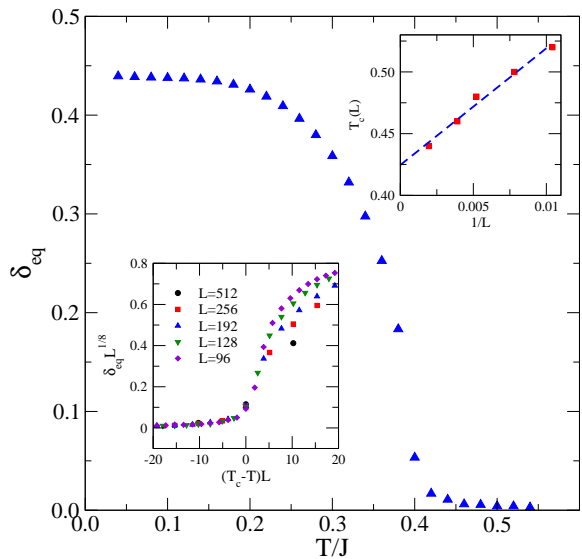}
\caption{Thermal melting of the SP phase. The main panel shows the 
evolution of the equilibrium bond distortion as the SP state melts to
a normal (paramagnetic) phase. $\delta_{eq}$ remains finite at
low temperatures, indicating a stable SP ground state. With increasing
temperature, the bond length distortion decreases and eventually vanishes 
via a continuous transition. The data is for a single system size
$N$=256x32 with parameters $\alpha=0.25$ and $\zeta=1.14$.  
The top right inset shows a plot of the $T_c(L) $ finite lattices
as a function of the inverse linear dimension to estimate $T_c$ in the
thermodynamic limit. The lower left inset shows the data for different
system sizes collapse to a single curve close to the transition temperature
for 2D Ising critical exponents ($\beta=1/8$ and $\nu=1$),
confirming the universality class of the transition.}
\label{fig:T1}
\end{figure}

In the final part of the work, we study the thermal phase transitions for the
ground state phases determined above. The N{\' e}el state in 2D is destroyed 
by any infinitesimal thermal fluctuations in accordance with the Mermin-Wagner
theorem, but the SP phase with a {\em discrete} broken symmetry persists to
finite temperatures. With increasing temperature, the equilibrium bond distortion
decreases and finally vanishes at a critical temperature via a thermal phase 
transition whose nature is probed in detail. We have extended the QMC studies 
to simulate the Hamiltonian (\ref{eqn:H}) at finite temperatures. As noted
in Section II, the strategy used to determine the ground state bond distortion fails
at finite temperatures because of the non-zero entropic term in the free energy. 
Instead, both the spin configurations and the bond distortions are evaluated 
using Monte Carlo updates. Since it
 breaks a two-fold discrete symmetry, the melting of the SP phase is expected 
to belong to the 2D Ising universality class. Fig.\ref{fig:T1} (main panel) shows 
the temperature dependence of the equilibrium bond length distortion, 
$\delta_{eq}$, for a single finite-size lattice. The data confirm that the distortion decreases 
monotonically with $T$ and eventually vanishes at a (size-dependent) critical 
temperature via a continuous phase transition. The estimate for the true
critical temperature in the thermodynamic limit is extracted from a finite-size 
scaling of the values for a wide range of finite-size systems. The universality class of
the thermal transition is verified by plotting $\delta_{eq}(t,L)L^{\beta/\nu}$ vs. $tL^{1/\nu}$, where
$t=(T_c-T)/T_c$ is the reduced temperature and $L$ is the system size. Close
to the transition temperature, the data for different system sizes are found to
collapse on a single curve when we use the known critical  exponents for the 2D
Ising universality class. The calculated specific heat (not shown here) is consistent with the
expected  2D Ising universality behavior, but the accuracy was found to be insufficient to 
extract the critical exponent.

\section{Summary}

A quantum Monte Carlo method has been used to study the $S=1/2$ antiferromagnetic
Heisenberg model on a square lattice with varying inter-chain interaction coupled to 
static bond phonons. Motivated by experimental observations in the inorganic 
quasi-1D spin-Peierls compound CuGeO$_3$,\cite{tranquada} the bond distortions are restricted 
to be only along the chains, and a single dimerization pattern, with a wave vector 
${\bf Q}=(\pi,\pi)$, is considered. It is found that in contrast to uncoupled chains, 
the transition to a dimerized spin-Peierls ground state occurs only when the 
spin-lattice coupling strength $\zeta$, exceeds a finite, non-zero critical value, $\zeta_c$,
at any non-zero inter-chain coupling $\alpha$. For $\zeta<\zeta_c$, the ground 
state has long range N{\'e}el order and zero spin gap, whereas for $\zeta>\zeta_c$, 
the ground state develops a finite dimerization accompanied by the opening up of a 
spin gap. The transition is found to be a discontinuous (first order) quantum phase 
transition. The value of the critical coupling depends on the strength of the inter-chain 
coupling and vanishes logarithmically as $\zeta_c \sim 1/\mbox{ln}\alpha$ 
as $\alpha \rightarrow 0$. The phase diagram in the parameter space of $\alpha$ 
and $\zeta$ is mapped out. Furthermore it is found that in the present model, the 
transition to the dimerized Peierls state is accompanied by the simultaneous vanishing 
of magnetic order, and there is no region of co-existing magnetic order and non-zero 
dimerization. Finally, the thermal transition of the dimerized state is studied in detail 
and is determined to belong to the 2D Ising universality class.


\begin{references}

\bibitem{pytte}
E. Pytte, Phys. Rev. B {\bf 10}, 4637 (1974).

\bibitem{cross}
M. C. Cross, and D. S. Fisher, Phys. Rev. B {\bf 19}, 402 (1979).

\bibitem{aws-sp}
A. W. Sandvik and D. K. Campbell, Phys. Rev. Lett. {\bf 83}, 195 (1999).

\bibitem{bursill}
R. J. Bursill, R. H. McKenzie and C. J. Hamer, Phys. Rev. Lett. {\bf 83}, 
408 (1999).

\bibitem{hase}
M. Hase, I. Terasaki, and K. Uchinokura, Phys. Rev. Lett. {\bf 70}, 3651 (1993).

\bibitem{riera}
J. Riera, and A. Dobry, Phys. Rev. B {\bf 51}, 16098 (1995).

\bibitem{nishi}
M. Nishi, O. Fujita, and J. Akimitsu, Phys. Rev. B {\bf 50}, 6508 (1994).

\bibitem{if1}
S. Inagaki, and H. Fukuyama, J. Phys. Soc. Jpn. {\bf 52}(10), 3620 (1983).

\bibitem{if2}
S. Inagaki, and H. Fukuyama, J. Phys. Soc. Jpn. {\bf 57}(2), 1435 (1988).

\bibitem{katoh}
N. Katoh, and M. Imada, J. Phys. Soc. Jpn. {\bf 62}, 3728 (1993); 
{\it ibid.} {\bf 63}, 4529 (1994).

\bibitem{koga}
A. Koga, S. Kumada, and N. Kawakami, J. Phys. Soc. Jpn. {\bf 68}, 642 (1998);
{\it ibid.} {\bf 68}, 2373 (1999).

\bibitem{al-omari}
 A. Al-Omari, J. Phys. Soc. Jpn. {\bf 69}(10), 3387 (2000).

\bibitem{sirker}
J. Sirker, A. Kl{\"um}per, and K. Hamacher, Phys. Rev. B {\bf 65}, 134409 (2002).

\bibitem{ono}
Y. Ono, and T. Hamano, J. Phys. Soc. Jpn. {\bf 69}(6), 1769 (2000).

\bibitem{tang} 
S. Tang, and J. E. Hirsch, Phys. Rev. B, {\bf 37}, 9546 (1988).

\bibitem{sumit} 
S. Mazumdar, Phys. Rev. B, {\bf 36}, 7190 (1987).

\bibitem{zhang} 
F. C. Zhang, and P. Prelov\v{s}ek, Phys. Rev. B, {\bf 37}, 1569 (1988).

\bibitem{khomskii}
M. Mostovoy, D. Khomskii, and J. Knoester, Phys. Rev. B {\bf 58}, 8190 (1998).

\bibitem{saito}
M. Saito, J. Phys. Soc. Jpn. {\bf 68}(9), 2898 (1999).

\bibitem{affleck}
E. S{\o}rensen, and I Affleck, D. Augier, and D. Poilblanc, Phys. Rev. B {\bf 58}, R14701 (1998).

\bibitem{dobry}
A. Dobry, P. Hansen, J. Riera, D. Augier, and D. Poilblanc, Phys. Rev. B {\bf 60}, 4065 (1999).

\bibitem{melin1}
M. Fabrizio, R. M{\'e}lin, and J. Souletie, Eur. Phys. J. B {\bf 10}, 607 (1999).

\bibitem{melin2}
R. M{\'e}lin, Eur. Phys. J. B {\bf 16}, 261 (2000); {\it ibid.} {\bf 18}, 263 (2000).

\bibitem{ji}
Y. Ji, J. Qi, J.-X. Li, and C.-D. Gong, J. Phys: Cond. Mat. {\bf 9}, 2259 (1997).

\bibitem{yuan}
Q. Yuan, Y. Zhang, and H. Chen, Phys. Rev. B {\bf 64}, 12414 (2001).

\bibitem{wang1}
Y. R. Wang, Phys. Rev. B, {\bf 43}, 3786 (1991).

\bibitem{wang2}
Y. R. Wang, Phys. Rev. B, {\bf 46}, 151 (1992).

\bibitem{azzouz}
M. Azzouz, and C. Bourbonnais, Phys. Rev. B {\bf 53}, 5090 (1996).

\bibitem{low}
C. H. Aits, and U. L{\"o}w, Phys. Rev. B {\bf 68}, 184416 (2003).

\bibitem{rts}
P. Sengupta, R. T. Scalettar, and R. R. P. Singh, Phys Rev. B {\bf 66}, 144420 (2002).

\bibitem{tranquada}
K. Hirota, D. E. Cox, J. E. Lorenzo, G. Shirane, J. M. Tranquada, M. Hase, K. Uchinokura, H. Kojima, Y. Shibuya, and I. Tanaka, Phys. Rev. Lett. {\bf 73}, 736 (1994).

\bibitem{metropolis}
N. Metropolis, A. W. Rosenbluth, M. N. Rosenbluth, A. H. Teller, and E. Teller, J. Chem. Phys. {\bf 21}, 1087 (1953).

\bibitem{miyashita}
H.~Onishi and S.~Miyashita, J. Phys. Soc. Jpn. {\bf 69}, 2634 (2000).

\bibitem{aws1}
A. W. Sandvik, J. Phys. A {\bf 25}, 3667 (1992).

\bibitem{aws2}
A. W. Sandvik, Phys. Rev. B {\bf 56}, 11678 (1997).

\bibitem{sseloop}
A. W. Sandvik, Phys. Rev. B {\bf 59}, 14157 (1999).

\bibitem{dirloops}
O. F. Sylju\r{a}sen and A. W. Sandvik, Phys. Rev. E {\bf 66}, 046701 (2002).

\bibitem{multichain}
A. W. Sandvik, Phys. Rev. Lett. {\bf 83}, 3069 (1999).

\bibitem{yasuda}
C. Yasuda, S. Todo, K. Hukushima, F. Alet, M. Keller, M. Troyer, and H. Takayama, Phys. Rev. Lett. {\bf 94}, 217201 (2005).

\bibitem{chn}
S. Chakravarty, B. I. Halperin and D. R. Nelson,  Phys. Rev. Lett. {\bf 60}, 1057 (1988);
 Phys. Rev. B {\bf 39}, 2344 (1989).

\bibitem{barber}
For a review of finite-size scaling, 
see: M. N. Barber in {\it Phase Transitions
and Critical Phenomena, Vol. 8}, ed. Domb and Lebowitz (Academic Press, 1983).

\bibitem{campostrini}
M. Campostrini, M. Hasenbusch, A. Pelissetto, P. Rossi, and E. Vicari, 
Phys. Rev. B {\bf 65}, 144520 (2002).

\end{references}
\end{document}